\documentclass[10pt,journal,compsoc]{IEEEtran}
\usepackage{pdfpages}
\usepackage{graphicx}
\usepackage{setspace}
\usepackage{blindtext}
\usepackage{color}
\usepackage[latin9]{inputenc}
\usepackage[T1]{fontenc}
\usepackage{url}
\usepackage{float}
\usepackage{breakurl}

\usepackage{comment}

\ifCLASSOPTIONcompsoc
  \usepackage[nocompress]{cite}
\else
  \usepackage{cite}
\fi
\ifCLASSINFOpdf
\else
\fi
\begin{document}
	
%
\title{Addressing multiple bit/symbol errors in DRAM subsystem}
%
%
%

\author{Ravikiran~Yeleswarapu
        and~Arun~K.~Somani 
\IEEEcompsocitemizethanks{\IEEEcompsocthanksitem Ravikiran Yeleswarapu and Arun K. Somani are with Department
of Electrical and Computer Engineering, Iowa State University, Ames,
IA, 50010.\protect\\
E-mail: mailravi@iastate.edu, arun@iastate.edu}}
\IEEEtitleabstractindextext{%
\begin{abstract}
As DRAM technology continues to evolve towards smaller feature sizes and increased densities, faults in DRAM subsystem are becoming more severe. Current servers mostly use CHIPKILL based schemes to tolerate up-to one/two symbol errors per DRAM beat. Multi-symbol errors arising due to faults in multiple data buses and chips may not be detected by these schemes. In this paper, we introduce Single Symbol Correction Multiple Symbol Detection (SSCMSD) - a novel error handling scheme to correct single-symbol errors and detect multi-symbol errors. Our scheme makes use of a hash in combination with Error Correcting Code (ECC) to avoid silent data corruptions (SDCs). SSCMSD can also enhance the capability of detecting errors in address bits.

$\-$ $\-$ $\-$ $\-$ We employ 32-bit CRC along with Reed-Solomon code (ECC) to implement SSCMSD for a x4 based DDRx system. Our simulations show that the proposed scheme effectively prevents SDCs in the presence of multiple symbol errors. We are able to achieve this improvement in reliability with similar READ latency as compared to existing ECC. For this design, we need 19 chips per rank (storage overhead of 18.75 percent), 76 data bus-lines and additional hash-logic at the memory controller.
\end{abstract}

}

\maketitle

\IEEEdisplaynontitleabstractindextext

%
\IEEEpeerreviewmaketitle

\IEEEraisesectionheading{\section{Introduction}\label{sec:introduction}}

%
%
%
%
\IEEEPARstart{F}{ailures} in DRAM subsystem are one of the major sources of crashes due to hardware errors in computing systems [2]. As DRAM technology continues to evolve towards smaller feature sizes and increased densities, faults in DRAM devices are predicted to be more severe. Small cell dimensions limit the charge that can be stored in them. This results in lower noise margins. As cell density increases, coupling (or crosstalk) effects come into picture. In-fact, researchers have recently identified "disturbance error" [19] in newer DRAM devices. This error has a cross-device correlation, hence will lead to multi-bit errors across different devices in a rank.

Each generation of DDRx family has doubled the transfer rates and reduced I/O voltages, and therefore, transmission errors in the Memory controller-DIMM interface are on the rise [1, 26]. Most of the servers use CHIPKILL [4] based reliability schemes. They, can tolerate only one or two symbol errors per beat. Multiple bit errors spread across the chip boundaries of a rank may not be detected by these schemes. Errors in bus along with growing device failures increase the frequency of multi-bit errors in the data fetched from DRAM subsystems.
 
Numerous field studies such as [1,3] studied large scale data-centers and predict that future exascale systems may require stronger reliability schemes than CHIPKILL. These studies base their analysis using limited protection mechanisms/logging capabilities and therefore the actual failure rates might be greater than their assessments. 

We first describe our error model, which captures the effect of various type of faults that may occur in DRAM devices and the data-bus. This model complements recent efforts such as AIECC [13], which focus on faults in address, command and control signals. We then propose a new error handling mechanism - Single Symbol Correction Multiple Symbol Detection (SSCMSD). As single symbol errors/beat are more frequent [3], our mechanism uses ECC to correct them. In addition we use a hash function to detect the less frequently occurring multi-bit (or symbol) errors. A hash function will detect multi-symbol errors with a high probability. It is the judicious combination of the two, i.e ECC and hash that makes our scheme effective.

We believe that SSCMSD is a very effective reliability mechanism for HPC/data-centers. More frequently occurring single symbol errors are corrected to achieve low recovery time. On the other hand, relatively infrequent, multi-symbol errors are detected by SSCMSD. Our scheme is also suitable for Selective Error Protection (SEP [31]), as we can enable/disable the enhanced detection capability provided by the hash for certain applications.

The rest of the paper is organized as follows. Section 2 introduces DDRx subsystem, Section 3 describes prior work in the field of memory reliability. In Section 4, we perform preliminary experiments to understand the impact of multi-symbol errors to Single Symbol Correcting Reed Solomon (SSC-RS) code  and to validate our simulation framework. Our error model is described in Section 5. Section 6 details our SSCMSD scheme. In Section 7, we evaluate our scheme with other mechanisms.  In Section 8, we summarize our work with Conclusion. 

\section{DDRx Memory Organization}

A DDRx [6,30] based memory is organized into hierarchical groups to enable designers to trade bandwidth, power, cost and latency while designing memory subsystems. At the topmost level, the subsystem comprises one or more channels. Each channel is made up of one or more DDRx DIMMs, a shared Data-bus, Address/Command bus, Control and Clock signals. Each DIMM includes multiple DRAM chips which are grouped into multiple "ranks". Typically, each DIMM has one, two, four or eight ranks. Furthermore, each chip has multiple independent banks. Each bank is composed of multiple sub-arrays [21] and a global sense amplifier. Each sub-array is further organized into a matrix of rows and columns with a sense amplifier. Figure 1 shows the organization of a channel which is composed of two x4 (transfer width - 4 bits) based DIMMs. The data bus is organized into sixteen groups or "lanes", each lane is shared by DRAM devices (or chips) across a channel. Address/command and control buses drive all the devices in the channel.

\begin{figure}[t]
	\includegraphics[scale=0.3]{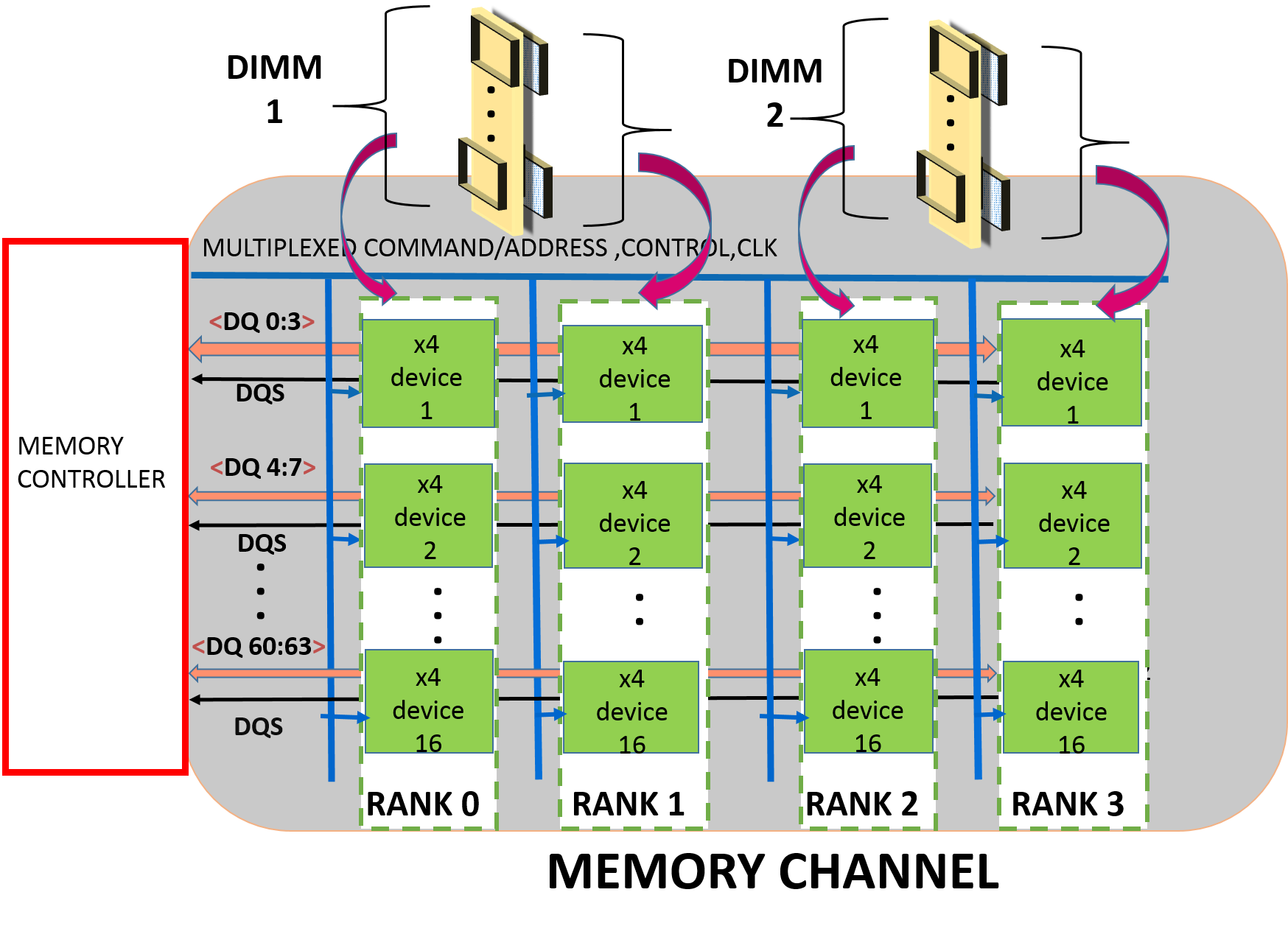}
	\caption{Memory channel - Memory controller is connected to DRAM modules (DIMMs) through shared bus.} 
\end{figure}

The memory controller (MC) handles memory requests from processor/cache/IO devices. As shown in Figure 1, the MC communicates address, commands, and data to the DRAM ranks over the channels. Typically, read/write cache miss require 64-byte data to be transferred between MC and DRAM memory subsystem. In this paper, we refer this 64-byte data (plus additional redundancy if any) as a cache-line. This is communicated in eight "beats" (8-half bus cycles). For a DRAM subsystem composed of DDR4, x4 devices, each beat activates an entire rank [33] (sixteen devices) and MC fetches/sends 64 bits of data per beat.

\section{PRIOR WORK}
This section summarizes schemes currently used by the industry and recent academic efforts to improve the reliability of DRAM subsystem. SECDED [7] and CHIPKILL [4] mechanisms were developed to address DRAM device errors. JEDEC introduced four schemes in DDR4 [14] to partially address signal integrity errors. MEMGUARD [23],  Bamboo-ECC [12] and AIECC[13] are recent academic efforts which are closely related to our work.  
\subsection{SECDED}

During 1990s, memory modules in servers were protected by using SECDED Codes. These codes make use of redundant (or check) bits to correct single-bit or detect double bit errors in a beat. For a typical beat size of 64 bits, SECDED code [7] makes use of eight redundant bits. SECDED design can correct 1-bit error or detect 2-bit errors in 64 bits (per beat) with 12.5\% redundancy and 8 additional bus lines/channel. In practice, it can detect/mis-correct some multi-bit errors [12] as well.

\subsection{CHIPKILL Correct}

As the demand for larger, high-density memory modules increased in the server industry, there was a need to protect against a single device failure. IBM introduced the "CHIPKILL Correct" error model to tolerate the failure of a single DRAM device in a rank.

CHIPKILL implementations make use of Reed Solomon (RS) Codes. RS codes use Galois "symbol" (set of bits) based arithmetic [8] and like SECDED use additional logic to generate codewords (set of data and check symbols) using data symbols. The circuit complexity of RS code increases with the symbol size. Therefore, small symbol sized RS codes such as 4-bit and 8-bit ones are more commonly used. There are two popular versions of chipkill:
\\\\
\textbf{1) SSCDSD (Single Symbol Correct, Double Symbol Detect) CHIPKILL}: AMD's 2007 design [10] and Sun UltraSPARC [11] provide SSCDSD capability for x4 DRAM devices by using 4-bit symbol RS code with four check symbols. To maintain redundancy at 12.5\%, this design uses 32 data symbols (128 bits), 4 check symbols (16 bits) per beat with 144-bit data bus and 36 devices per rank. The nature of the design is such that it "over fetches", i.e. two cache lines are accessed during a memory transaction (8 beats * 32 data devices/rank * 4 bits = 128 Bytes) and uses 144 bit data bus. Therefore, it may result in increased energy consumption. 
\\\\
\textbf{2) SSC (Single Symbol Correction) CHIPKILL}: To reduce cache access granularity, in 2013, AMD developed a SSC based 8-bit symbol RS code [5] for x4 DRAM devices. This scheme uses the 72 bit data bus and 18 devices per rank (64 data + 8 redundant bits/beat). In this design, bits from two successive beats are interleaved to form one codeword  with "Chipkill" functionality [5, 40]. For 8 beats from x4 devices, each cache request makes use of four codewords. Each codeword comprises 16 data and two check symbols with a redundancy of 12.5\%. This design is used as our baseline for comparison.

When there are >1 symbol errors/codeword (mostly due to multiple chip failures), AMD uses history based hardware-software approach to cover these scenarios [5].


\subsection{DDR4 Bus reliability mechanisms}

\textbf{1) WRITECRC}: WRITECRC is designed to detect transmission errors in data during WRITE operation. In this design, the memory controller generates an 8-bit CRC checksum for the entire write data burst (8 beats) to each chip/data-lane [14] of the rank. These 8 bits are sent over two additional beats after the data is sent to the individual chips. Each DRAM chip also has logic to re-compute the CRC checksum and compare it with checksum sent by the controller. Such a design allows the chips to detect errors before storing them and provides an option to retry the transmission of the data. However, transmission errors during READs (not covered by WRITECRC) may also lead to SDCs with the baseline scheme.
\\\\	
\textbf{2) CA (Command/Address) parity}: CA parity uses an additional pin (or bus-line) to transfer even parity of the CMD/ADD signals to every DRAM chip. It cannot detect an even number of bit-errors on the CMD/ADD signals.
\\\\
\textbf{3) Data Bus Inversion (DBI)}: DBI is designed to protect against Simultaneously Switching Noise (SSO) [20] during data transmission. This scheme is only available for x8, x16 DDR4 chips. With 8 Data bits/pins and an additional 9th pin per each data-lane, DBI ensures that at least 5 out of 9 pins are "1"s. This avoids the situation where all bits go from 0 to 1 or from 1 to 0 to improve the signal integrity of data bus.
\\\\
\textbf{4) Gear Down Mode}: Gear-down mode allows the MC to lower transmission rate of command/Address and control signals to trade-off latency and command bandwidth for signal quality while maintaining high data rates/bandwidth.

\subsection{Memguard [23]}
Memguard is a reliability scheme designed to detect multi-bit errors in DRAMs without using redundant storage. It makes use of two registers (READHASH, WRITEHASH) and custom logic at the memory controller (MC). Whenever there is a memory transaction between the last level cache and the DRAM, the logic at MC computes a hash value for this transaction and READHASH/WRITEHASH registers are updated. This scheme does not store the hash values in the memory as they use incremental multi-set hashing technique [35]. By periodically synchronizing the two hash registers at the MC, this scheme detects errors in data. Memguard relies on OS-checkpointing for error recovery. 

Although this scheme can detect multi-bit (or multi-symbol) errors, on its own it is not suitable for HPC/datacenters due to the high recovery time associated with checkpointing and synchronization. Also, Memguard is effective only against soft errors. Although our design is motivated by Memguard's scheme, we do not use incremental multi-set hashing technique which is at the core of Memguard's design and instead store hash along with data and ECC bits in the DRAM (use redundancy). Thus, unlike Memguard, we employ ECC and hash to provide correction/detection capability for each cacheline and do not require any synchronization for error detection. This ensures faster recovery, effectiveness against both permanent and soft errors, and is therefore suitable for  HPC/datacenters/servers.




\subsection{QPC Bamboo ECC [12]}
QPC Bamboo-ECC is an 8-bit symbol RS based scheme designed to target more frequently occurring error patterns. They provide CHIPKILL capability with 12.5 \% redundancy for x4 based memory systems and show that they perform better than AMD's RS-SSC CHIPKILL in reducing SDCs for certain type of faults. Since they use one codeword for the entire cache-line, their design leads to increased READ latency. 

Our goal in this paper is to consider more realistic error model based on the nature of faults and develop an appropriate scheme to protect against them. We demonstrate that Bamboo-ECC and extended Bamboo-ECC (same overhead as ours) are prone to Silent Data Corruptions when faults are spread across multiple chips/buses. 

\subsection{AIECC-All Inclusive ECC [13]}
AIECC is a suite of mechanisms designed to protect against CCCA (clock, control, command, address) faults without additional redundant storage or new signals to and from memory. 

Our work is orthogonal to AIECC. We focus on improving reliability against device, bus errors while AIECC focuses on CCCA errors. The reliability of future memory systems can be improved by incorporating our scheme along with AIECC.

\section{Preliminary Experiments}
In the presence of an error, a generic reliability scheme reports it as either a Correctable Error (CE) or a Detectable but Uncorrectable Error (DUE). When an error is outside of the coverage of the scheme, it can result in Detectable but Miss-corrected Error (DME) or Undetectable Error (UE). DMEs and UEs are collectively called as Silent Data Corruptions (SDCs) as the scheme inadvertently forwards corrupt data without raising an alarm.

The baseline scheme uses RS (18, 16, 8) systematic SSC code.  A RS (n, k, m) codeword has k data symbols and n-k check symbols with m bits per symbol. The minimum hamming distance between any two codewords is n-k+1 (3 in this case). It can correct $\lfloor(n-k)/2\rfloor$ (1 for baseline scheme) symbol errors. When there is an error across multiple symbols of a codeword, this RS decoder can either identify it to be uncorrectable error (DUE) or "miss-correct" it to another codeword thinking it to be a single symbol error of another codeword (DME) or fail to detect presence of the error (UE). This can result in Silent Data corruptions in the baseline scheme. We therefore devised simple set of experiments to assess the amount of Silent Data corruptions in the baseline in presence of multi-symbol errors.


We developed an in-house simulator to perform our experiments. We used open source software [16], [36], [41] to develop galios (symbol-based) arithmetic, RS encoder, decoder. We use generator polynomial - $G(x) = (x-a^1)(x-a^2)...(x-a^N)$ (where N - number of ECC symbols/CW) to construct RS code. Our decoder uses Berlekamp Massey algorithm for correcting/detecting errors. Due to the simplicity of hardware design, most of the hardware implementations use either algorithm based on Euclidean approach or Berlekamp Massey to implement Reed Solomon decoder. We also verified that the simulation results are similar with Euclidean based RS decoder.


For each iteration of the experiment, we fed random 16 byte dataword to RS encoder and stored the 18-symbol codeword in an array. With the help of 18-symbol error mask, we inserted errors into the stored codeword.
We then, decoded the stored codeword (with errors) using the RS-decoder. The decoder flagged whether each codeword had
"No Errors" or "Detectable but Uncorrectable Errors" or "Correctable
Errors". If the decoder detected a correctable error in a codeword,
it corrected the corresponding stored-codeword.
Next, we retrieved the stored
data word processed by RS-decoder and compared it with the original data word to identify
silent data corruptions. 

We executed three experiments - introducing random 2, 3 and 4 symbol errors
per codeword. Each of this experiment
was run for ten iterations and each iteration handled 1 billion random datawords. Table 2 lists the mean \%  across 10 iterations for the number of miscorrections, detected but uncorrectable
errors and undetected errors. The standard deviation for each of the experiments (except for undetected errors with random 2 symbol errors) was up to 13,000.  We also show the mean of undetected errors along with mean \% to show give a glimpse of actual number of undetected errors we encountered.     
\begin{table}
	\begin{tabular}{|c|c|c|c|}
		\hline 
		Experiments & Miscorrected & Detected but  & Undetected \\&&Uncorrected&  \tabularnewline
		\hline 
		\hline 
		2 Symbol  & 6.3\% & 93.7\% & 0\% (0) \\Errors/CW&&&  \tabularnewline
		\hline
		3 Symbol  & 6.9\% & 93.1\% & -> 0\% \\Errors/CW&&&(\~10,000)   \tabularnewline
		\hline
		4 Symbol  & 7.0\% & 93\% & -> 0\% \\Errors/CW&&&(\~10,000)  \tabularnewline
		\hline
		
	\end{tabular}
	
	\caption{Experiments/Results of Random multi-symbol data errors for RS (18,16,8).}
\end{table}
\subsection{Analytical Analysis}

We can explain these results of our experiments with the help of analytical methods. Figure 2 depicts the codespace of the baseline RS (18, 16, 8) code. In the figure, stars represent valid codewords and diamonds represent non-codewords. Due to errors, a particular codeword (say CW1) gets corrupted and may be detected by RS decoder as a non-codeword (diamond) or as other codeword in the space [12]. The dotted hypersphere which is  HD = 1 away from codeword represents the correction range of the SSC. All the words on this sphere will be corrected to the codeword on the center of the sphere (in this case CW1). Words on HD=2 hypersphere (solid line in green) are either detected as errors or miscorrected to the adjacent codeword. Words on the dashed sphere (HD=3) are either correctly detected as errors or undetected (falsely detected as adjacent codeword) or miscorrected as another codeword.

\begin{figure}
	\centering
	\includegraphics[scale=0.32]{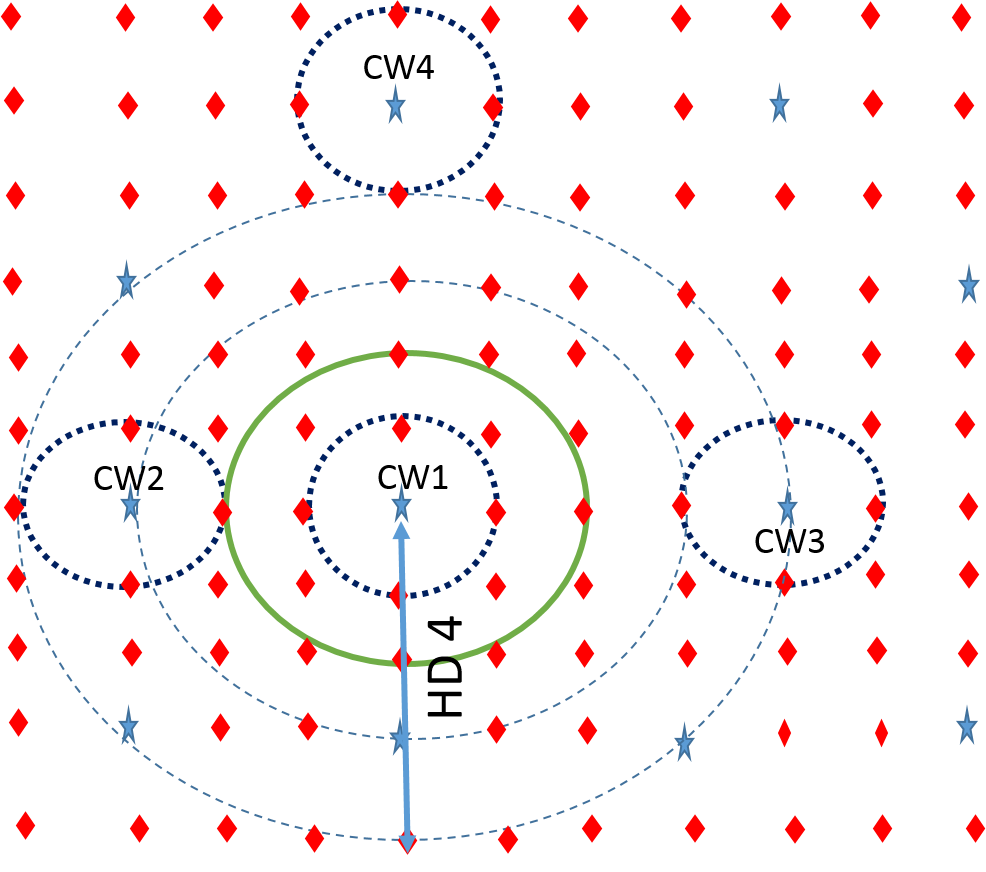}
	\caption{N tuple space representation of Reed Solomon SSC code. Stars are codewords which are atleast 3 Hamming Distance apart.}
	
\end{figure}
For a generic RS (n, k, m) code, the total n-tuple space available is $2^{n*m}$. Out of this space, the number of codewords are $2^{k*m}$. Assuming that the space is uniformly distributed among the codewords, we can say that the space around (or owned by) each codeword is $2^{n*m} / 2^{k*m} $.

If we introduce "e" symbol errors from a given codeword (say CW1), all such words lie on a hypersphere at HD=e from the codeword. If "e" is greater than minimum HD of "n-k+1", this sphere may also contain other codewords. For example, as shown in Figure 2, the RS code has two codewords (CW1 and CW2) which are HD=3 apart. If we introduce 4 symbol errors from CW1, the hypersphere centered on CW1 with radius 4 also contains CW2. On an average, the number of such codewords $C_{e}$ on or inside a hypersphere HD=e away is approximately given by dividing the total number of words inside the sphere by number of words "owned" by each codeword : 

\begin{equation}
C_{e} = \frac{\sum _{\alpha =1}^{e}{n}_{{C}_{\alpha }\left({2}^{m}-1\right)}\alpha }{{\left({2}^{m}\right)}^{n-k}} - 1
\end{equation}

\begin{figure}
	\centering
	\includegraphics[scale=0.37]{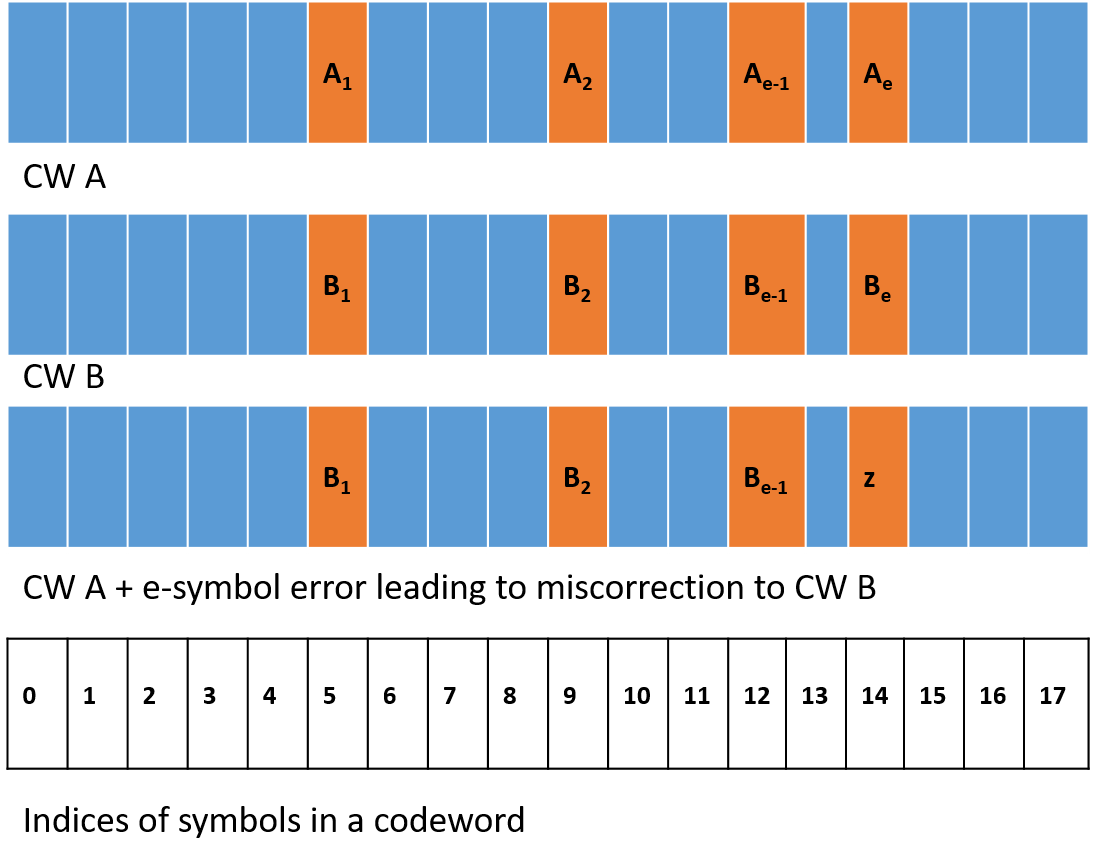}
	\caption{CW A, CW B are two 18-symbol codewords which are HD = e apart. Notice that dataword "CW A + e-symbol error" is HD = 1 (differ in 1 symbol) from CW B.}
	
\end{figure}

The RS decoder "mis-corrects" such an "e" (where $e > (n-k+1)$) symbol error from a given codeword when the e-symbol error also  :
1)	falls on HD = 1 sphere of another CW which is HD = e away OR
2)	falls on HD = 1 sphere of another CW which is HD = e-1 away OR 
3)	falls on HD = 1 sphere of another CW which is HD = e+1 away. For example, Figure 2 shows a hypersphere at HD = 4 away from CW1. This sphere represents all the 4-symbol errors from CW1. Few words on this sphere get mis-corrected to CW3, which is at HD = 4 away from CW1. Due to presence of CW2 at HD = 3 away from CW1, few other words on sphere HD=4 also fall on HD=1 sphere of CW2 and therefore get mis-corrected. Similarly, few other words on this HD = 4 sphere also fall on HD = 1 sphere of CW4 which is at HD = 5 away from CW1.

Using eq (1), the number of CWs at HD = e from a given codeword is given by $C_{e} - C_{e-1}$ which is equal to

\begin{equation}
\frac{{n}_{{C}_{e}\left({2}^{m}-1\right)}e}{{\left({2}^{m}\right)}^{n-k}}
\end{equation}

Now, due to the presence of one CW at HD = e from a given CW (say CW A), more than one "e" symbol errors are mis-corrected. Figure 3 shows two codewords CW A and CW B which are HD=e away. There will be exactly  ${e}_{{C}_{e-1}\cdot \left({2}^{m}-2\right)}$ number of "e" symbol errors from CW A which are HD=1 away from CW B and hence will be miscorrected to CW B. Combining this with eq (2) we get the expression for total number of "e" symbol errors from a given codeword CWA that will be miscorrected due to presence of codewords at HD=e from CW A:

\begin{equation}
m_{e} = \frac{{n}_{{C}_{e}}\cdot {\left({2}^{m}-1\right)}^{e}\cdot {e}_{{C}_{e-1}}\cdot \left({2}^{m}-2\right)}{{\left({2}^{m}\right)}^{n-k}}
\end{equation}
Similarly we can calculate number of "e" symbol errors that will be miscorrected due to presence of codewords at HD = e-1, HD = e+1 given by (4) and (5) respectively :

\begin{equation}
m_{e-1} = \frac{{n}_{{C}_{e-1}}\cdot {\left({2}^{m}-1\right)}^{e-1}\cdot {\left(n-e+1\right)}_{{C}_{1}}\cdot \left({2}^{m}-2\right)}{{\left({2}^{m}\right)}^{n-k}}
\end{equation}

\begin{equation}
m_{e+1} =\frac{{n}_{{C}_{e+1}}\cdot {\left({2}^{m}-1\right)}^{e+1}\cdot {\left(e+1\right)}_{{C}_{e}}}{{\left({2}^{m}\right)}^{n-k}}
\end{equation}

The total number of "e" symbol errors from a CW is given by ${n}_{{C}_{e}\cdot \left({2}^{m}-1\right)}e$. Therefore, the fraction of miscorrections in the total set of "e" symbol errors from a CW is given by :

\begin{equation}
m_{total} = \frac{1}{{n}_{{C}_{e}\cdot \left({2}^{m}-1\right)}e}\cdot \left( m_{e} + m_{e-1}+ m_{e+1}\right)
\end{equation}

Using (6), we calculate the fraction of mis-corrections for the experiments. For the first experiment (Random 2 Symbol errors) as RS (18, 16, 8) code has a minimum HD of 3, there are no codewords at $HD = 2$ or $HD = 1$, therefore $m_{e}$ and $m_{e-1}$ do not contribute to the expression in eq (6). We calculate the fraction of miscorrections for this experiment to be 6.3 \%. This value corroborates with the experimental results shown in Table 1. The total information space available for single symbol correcting RS (18,16,8) is $2^{18*8} (2^{n*m})$. Out of this, $2^{16*8}(2^{k*m})$ are to be used as codewords. As the fraction of codewords over the total space is only  $2^{-16} (2^{16*8} / 2^{18*8})$, as the code-space is sparsely populated 93.7 \% of random errors on HD=2 sphere do not fall on HD=1 spheres of other codewords. Also, as expected we do not observe any undetected errors in this experiment as there are no codewords at HD = 2. Similarly, we calculate the fraction of mis-corrections for the second and third experiments and find that these also corroborate with the experimental results in Table 1. 

As we are able to corroborate the experiment results with our analytical model, we have confidence that our experimental framework is able to accurately simulate Reed Solomon decoder and random error injection. Also, these results further motivated us to develop a solution to tackle SDCs in current/future DRAM subsystems.

\begin{table}[h]
	\begin{tabular}{|c|c|c|}
		\hline 
		Fault mode & Source & Error pattern \\&& per cacheline  \tabularnewline
		\hline 
		\hline
		1 bit fault & Particle strikes OR & 1 bit error/ \\&Cell failure in a  & cacheline\\&sub-array&\tabularnewline
		\hline 
		1 pin fault & Fault in 1DQ of  & 1 pin stuck at 0 \\& a bus lane OR &or 1 for all beats \\ &sub-array failure&\tabularnewline
		\hline
		Row/Chip/ & Failure of sub-array & 1 word related   \\Bank fault& row drivers/address  & to faulty chip  \\&decoding circuit&in all beats \\&&is stuck \\&&at 1 or 0\tabularnewline
		\hline
		Column  & Failure of single  & 1 bit stuck at 0/1\\fault& column in & in a cacheline\\ &a sub array&\tabularnewline
		\hline
		Bus fault & Fault in 1 bus lane &  errors in random  \\&&beat positions \\&&of a bus\tabularnewline
		\hline
		Correlated  & External noise or &Bus faults in \\Bus fault&coupling between  & consecutive\\& two consecutive &bus-lanes\\&bus-lanes&\tabularnewline
		\hline
		1bit/pin +  & Combine 1 bit/pin  & faults which \\other faults& fault with pin/&  lead  to \\&row/chip/bus & 2 symbol errors \\ &fault& \tabularnewline
		\hline
		Chip + Chip & Failure of 2 different  & 2 specific words   \\fault&chips&in all beats stuck \\&& at 0 or 1\tabularnewline
		\hline
		3 fault  & combine 3 of the & Random errors \\mode& above mentioned &in 3 words \\&faults& in all beats\tabularnewline
		\hline
	\end{tabular}
	
	\caption{Error Model }
\end{table}

\begin{table*}[t]
	\centering
	\begin{tabular}{|c|c|c|}
		\hline 
		Transmission Fault & Description / Cause  & Impact on Signal Integrity \tabularnewline
		\hline
		\hline 
		Dielectric Loss & Signals attenuate as a function of trace length and frequency & All data bits are affected, results  \\ &&in signal attenuation\tabularnewline 
		\hline 
		Skin effect & Resistance of conductor varies non-uniformly with frequency & All data bits are affected, results \\&&in signal attenuation\tabularnewline 
		\hline
		Electromagnetic & Electromagnetic/capacitive coupling of closely packed lines & few bus lines/lanes are affected  \\ interference& & at one point of time    \tabularnewline 
		 		
		\hline 
			   Skew & Path length variations result in timing variations & Random \tabularnewline 
		\hline
		       Jitter &Fluctuations in voltage, temperature and crosstalk between  & Difficult to model/characterize  \\&successive cycles of a given signal impact the propa-& \\ &gation time of the signal& \tabularnewline 
		 \hline
		  Inter symbol & Past signals on a line have residual effects on subsequent & No. of bit lines affected: Random, 	\\interference&signals of the same line &Data dependent\tabularnewline 
		  \hline 
		  	Simultaneously  & When many signals in a bus-lane switch, they induce coupling   & Data dependent. \\switching output &on other signals & \tabularnewline 
		  \hline      
	\end{tabular} 
	\caption{Summary of Data Transmission faults }
\end{table*}
%



\section{Error Model}

To represent the possible fault modes that may occur in current/future DRAM systems, we first describe our error model. This model covers various type of faults that arise in DRAM devices and the data-bus.

Faults in DRAM subsystems are caused due to a variety of sources such as cosmic rays [29], circuit failure, signal integrity etc.. These faults can be broadly categorized as transient or permanent. Transient phenomena corrupt memory locations temporarily, once rewritten these locations are free from errors. Permanent faults cause the memory locations to consistently return erroneous values.  

Field studies [1,3]  help us in understanding the trends of errors in DRAM subsystem up to a certain extent. We use this information along with nature of faults in DRAM subsystem to develop our error model (Table 2). Here, we describe the sources of these faults and the corresponding errors perceived per cache-line due to a particular fault type. Single bit faults are mainly due to failures in DRAM cells. Due to failure in a sub-array or one DQ pin (one bus line in a bus-lane), bits transferred over a single DQ pin are corrupted. Failures in circuitry inside chips such as sense amplifiers, address decoders etc. cause particular rows/columns/banks/chips to malfunction. For example, if a local row buffer (sense-amplifier) in a bank of a chip is stuck at 1, then all the bits fetched from the chip of particular READ request are read as "1". Therefore each word (bits provided by a chip in one beat) fetched from this chip will have all 1's for this particular READ.

Bus faults are another source of errors. According to 1st order analysis, bus lines act as a low pass filter. Since digital signals are composed of numerous frequencies, distinct components of this signals experience attenuation to a different degree giving rise to signal degradation. Reflection is another first order effect which results in signal degradation. Table 3 describes other sources of transmission faults [26] and their impact on signal integrity of the data bus. As most of the errors associated with bus faults are data-dependent or random, we expect random errors in different beats of a faulty bus. To simulate this behavior for a single bus fault, we use a random number to identify the erroneous beat positions among eight beats. We then inject random errors in these positions. We also consider correlated bus fault due to presence of external noise or coupling between two bus lanes. In this fault-mode, we expect two consecutive bus lanes to faulty. Similar to single bus fault, we first identify erroneous beat positions and inject random errors for these two consecutive bus lanes.  

We combine single bit faults (as they occur with higher frequency [28]) with other fault types to model 2-symbol/chip errors per codeword. With increase in the possibilities of fault occurrence especially in exascale systems, there is a higher possibility for other faults to occur simultaneously across three different chips/bus lanes. To cover such scenarios, we include 3-fault mode (fault which leads to random errors in three random chips/bus-lanes). 




\section{SSCMSD - A Novel Architectural Solution for multi-bit/multiple symbol errors}
We first carried out a set of experiments detailed in our error model to study the behavior of the baseline (SSC-RS (18,16,8)) scheme. The results are shown in the 2nd column of Table 5 in the Evaluation section. As described earlier, as the code-space is sparsely populated this scheme can detect many multi-symbol errors as well. However, as shown by the experiment results, the baseline is still prone to SDCs with multiple device and bus faults. Inspired by this observation, we chose to further decrease this SDC rate by improving the ECC scheme at the memory controller with minimal increase in redundancy (1 more redundant chip and 1 more bus lane). 

With one more chip and bus lane at our disposal, one can simply extend the baseline scheme. This extended baseline scheme uses three check symbols (instead of two used in baseline) per each codeword to provide SSC capability. The fourth column of Table 5 shows the performance of this scheme with our error model. This scheme has lower SDC rate when compared to baseline, but it is still prone to SDCs with multiple symbol errors.


An interesting point to note from these results is that the SDC rate is dependent on the type of error pattern a fault generates rather than on the number of bits/symbols being corrupted. For example, 1 bit + Chip fault corrupts 9 bits per CW and has 6\% SDC rate while 1-bit fault + 1-pin fault corrupts 3 bits of a particular CW and has a SDC rate of 7.6\% for the baseline scheme. Although we do not show the breakdown of SDCs into UEs and DMEs for the baseline in Table 5, our evaluation shows that for all the experiments of baseline and Extended-baseline schemes, SDCs occur mostly (99\%) due to miss-corrections (DMEs) from the SSC-RS decoder. Therefore, the stored information is subjected to errors from faults and due to errors induced by the decoder. These observations inspired us to use a hash function, as the hash value allows us to identify such arbitrary corruption in the data. Taking a cue from the design of Memguard [23], we use a non-cryptographic hash function to compute a signature of the data. We use this signature to detect multi-bit errors with high probability. By combining hash and CHIPKILL, we develop our new error handling scheme, called Single Symbol Correct, Multiple Symbol Detect (SSCMSD) CHIPKILL.

As shown in Figure 4, during WRITE operation, we can combine the hash and baseline CHIPKILL scheme in three possible ways : 
\\
\textbf{Scheme A}: Compute the hash of data and then use SSC encoder to encode data and hash. 
\\
\textbf{Scheme B}: Encode the data and then compute the hash of encoded data. 
\\
\textbf{Scheme C}: Encode the data and compute the hash of data in parallel.

As shown in Table 5 the baseline and extended baseline provide CHIPKILL(SSC) correction capability, but with multiple symbol errors, they result up-to 8\% SDC rate. The purpose of using the hash is to further reduce this SDC rate without impacting the existing reliability provided by SSC code. Therefore, while retrieving the data from the DRAM (READ operation), we use a simple, straight forward design to build upon the existing SCC capabilities. First, we perform the SSC decoding, in this process the decoder will tag each retrieved codeword to have NO Error OR Correctable Error OR Un-correctable Error. We then use the hash to validate the findings of the decoder. 

On analyzing Scheme B and Scheme C with this simple retrieval mechanism we find that there is a possibility of a false positive i.e  report data which was correctly handled by SSC decoder to be erroneous. This happens when the hash gets corrupted (erroneous). In this scenario, when there is a single symbol error or no error in the data/ECC symbols of a codeword, the decoder corrects it or reports that the retrieved data is free from errors, respectively. But, as the hash is corrupted in this scenario, the second step of the retrieval process reports that the data is erroneous. With Scheme A there is no scope for such false positives as hash is also correctable by SSC decoder. At the minimum, Scheme A guarantees to provide the reliability already offered by baseline (SSC decoder). In addition, it also provides capability to detect miscorrections OR undetected errors missed by the SSC decoder. Hence we identify Scheme A to be most suitable for our purpose.

As described earlier, Scheme A generates the hash of the data before encoding it with the RS-SSC encoder. This encoded data, hash pair (codeword) is stored in the memory during WRITE. When this stored codeword is retrieved from the memory during READ, we first employ the RS-SSC decoder to correct/detect errors. The RS-SSC decoder corrects up to one symbol error in each codeword to retrieve data, hash pair. As noted earlier, there is a possibility of silent data corruption in the retrieved data, hash pair if there are multiple symbol errors in the codeword. To detect this scenario, we recompute the hash of data retrieved from the SSC-RS decoder and compare it with the retrieved hash. If the hashes match, then with a high probability, we can conclude that there are no SDCs in the retrieved data. When the two hash values do not match, this indicates the presence of multiple symbol errors. Thus, we can effectively avoid silent data corruptions.

When there is up to one symbol error per codeword, this combined scheme (Scheme A, SCC decoding + Hash validation) corrects the codeword (similar to the baseline scheme) and pass on the requested 64-byte cache-line to the processor. Hence, applications waiting for this cache-line can resume their execution almost immediately on the processor. But if there is a multi-symbol error  in any of the codeword, our scheme would detect that with high probability and prevent silent data corruption. This is an improvement over the baseline scheme.


\begin{figure}
	\center	
	\includegraphics[scale=0.34]{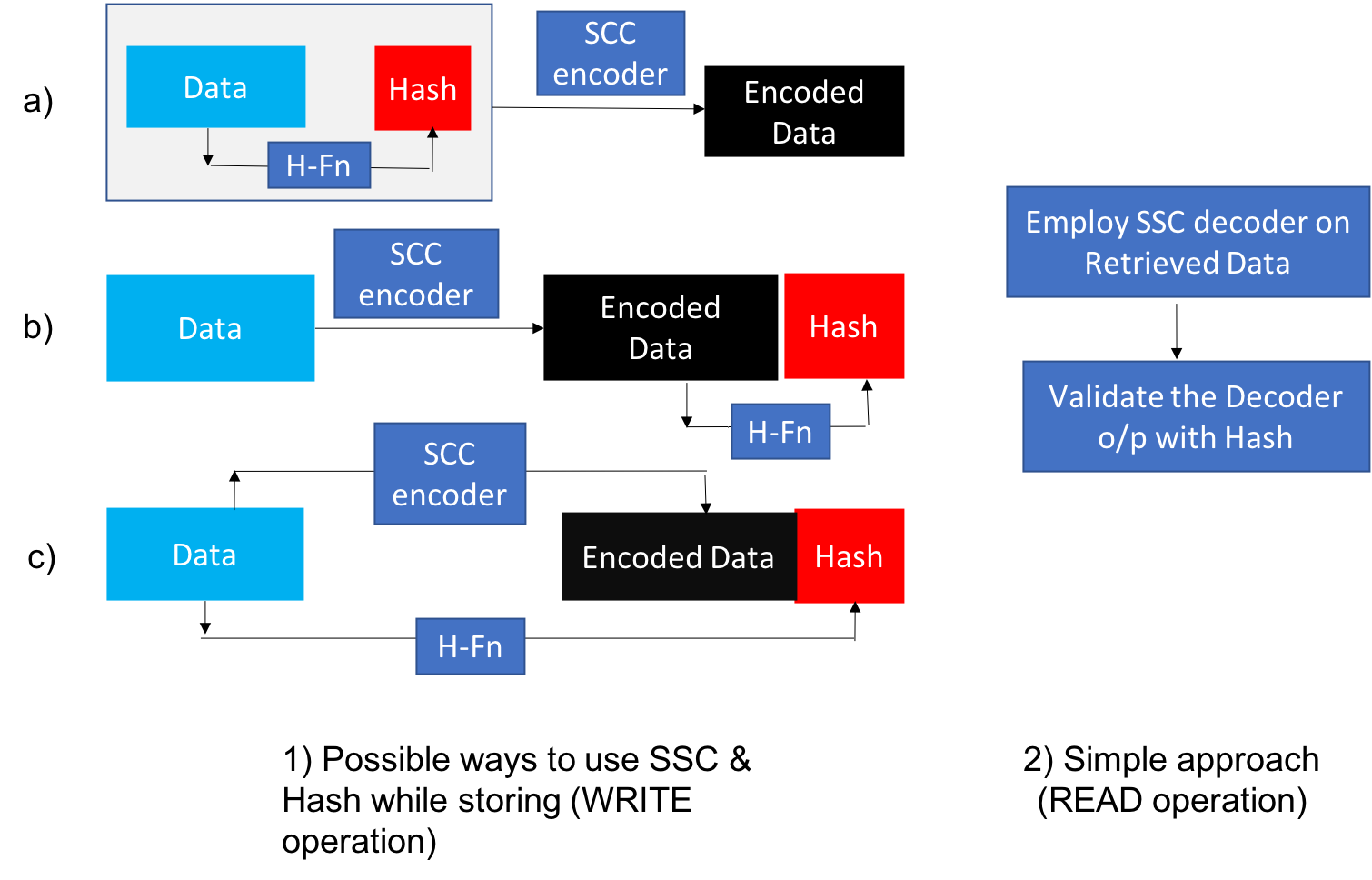}\caption{Possible hash and SSC combinations}
	
\end{figure}



\subsection{WRITE Operation}
As shown in Figure 5, during a WRITE operation, we use a hash function to generate 32 bit output (4 symbols) from the entire cacheline (64 Bytes). Similar to the baseline SSC-RS scheme, the 64 Byte data is divided into 4 blocks (Block$_{0}$-Block$_{3}$), each block is composed of 16 symbols. We  distribute the 4 symbol hash output across the 4 data blocks by combining each data block of size 16 data symbols with 1 hash symbol to obtain a dataword. The size of our "extended" dataword is 17 symbols, as opposed to 16-symbol dataword used in the baseline design. Each dataword is encoded using RS (19, 17, 8) code to obtain a 19-symbol codeword. This 19-symbol codeword is interleaved across 2 beats as in the baseline design. Therefore, we need a total of three additional chips (storage overhead of 18.75\%) per rank and 12 redundant bus-lines in every channel.

\begin{figure}
	\centering
	\includegraphics[scale=0.35]{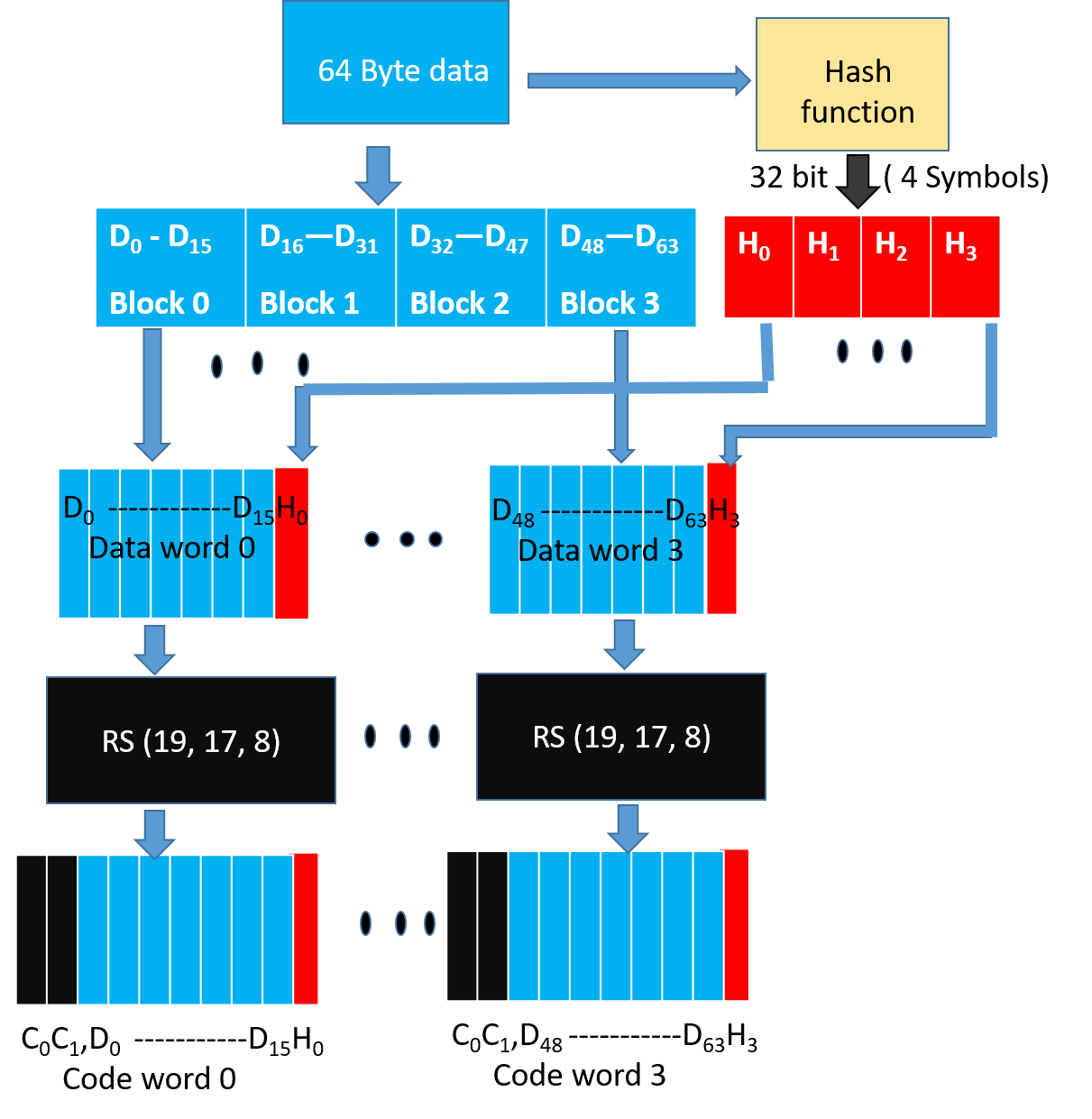}\caption{SSCMSD Design. 32 bit hash is split into 4 symbols (red). Each hash symbol is combined with 16 symbol data block (blue) to form a dataword.}
	
\end{figure}

\subsection{Read Operation}
Similar to the baseline scheme, during a READ request (or MISS) two consecutive incoming data beats at the memory controller are combined to obtain a 19-symbol codeword. As shown in Figure 6, for DDRx systems, the codewords of this READ request are obtained in four consecutive bus cycles. We need to employ SSC decoder on each codeword to obtain the 64-byte data and then validate this data with the help of hash function. As this two-step approach introduces additional latency to the READ MISS, in the following paragraphs, we describe our novel design to minimize this latency.

The SSC-Reed Solomon decoding on the received codewords is typically done in two phases. In the first phase, syndrome is computed to identify if there are any errors. Error correction (second phase) is computationally more expensive and therefore is triggered only when syndrome computation block detects errors. Since errors are relatively rare, the average delay incurred due to decoding will be close to the error free case where only the syndrome computation is performed. Study [34] mentions that delay of SSC-RS syndrome calculation is about 0.48 ns with 45nm VLSI design library. For DDR4 [4] with a memory clock frequency of 1200 Mhz, syndrome computation can be implemented within one memory cycle. 

The detection capability of our scheme depends on hash function properties such as length, collision resistance, avalanche behavior, distribution etc. [24]. Also, a non-cryptographic hash is suitable for our design as cryptographic hash functions are more complex in terms of computation time, which increases the memory latency.  Studies [24,32] show that non-cryptographic hashes - CityHash, MurmurHash, Lookup3 and SpookyHash have good properties with respect to avalanche behavior, collision resistance and even distribution. CRC-Hash is also widely used due to its simple hardware design and due to its linear property. We analyzed the hardware design of Lookup3, Spookyhash [25, 40], CRC-Hash [37] and found that these can be implemented using combination logic. Therefore these hash functions can be easily implemented within four memory cycles.
\begin{figure*}
	\center
	\includegraphics[height = 5.5 cm,width=\textwidth]{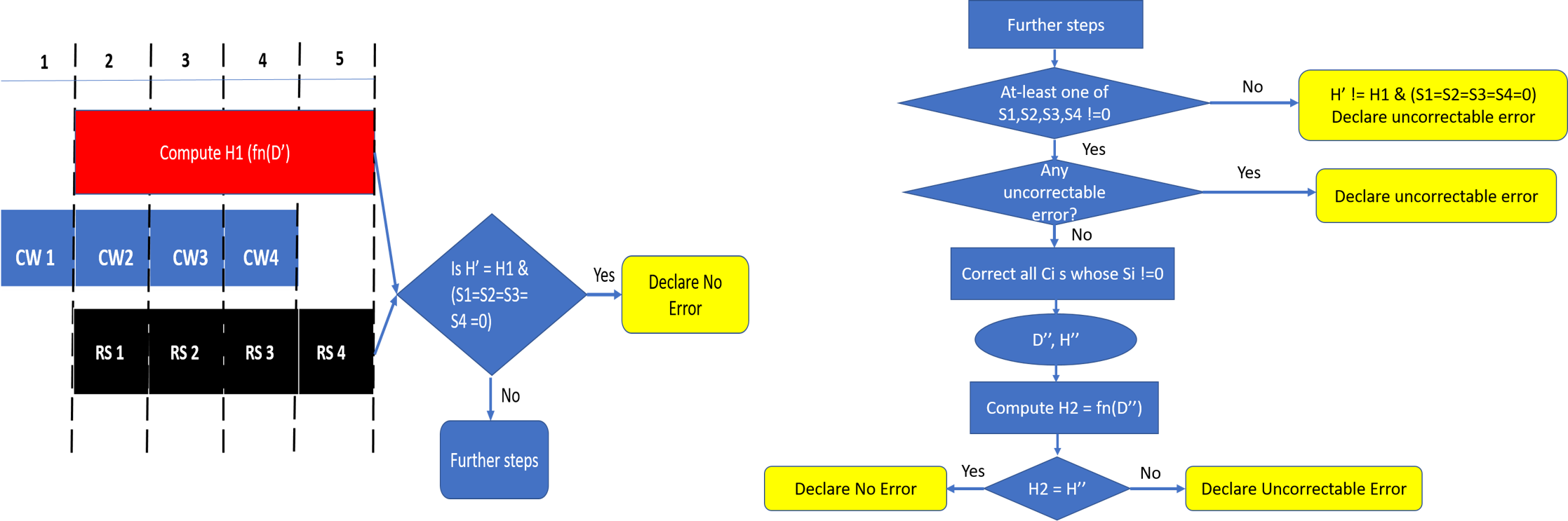}
	
	\caption{SSCMSD Design. During Read operation, syndrome computation and hash calculation are done in parallel. If any of the syndromes are non-zero AND if there are no uncorrectable errors, SSC correction is employed and Hash is computed to detect Silent Data Corruptions.}
	
\end{figure*}

Since we are using systematic SSC-Reed Solomon code, RS syndrome calculation and hash computation can be done in parallel. As shown in Figure 6, DDRx provides two beats of data per memory clock cycle, hence SSC-RS syndrome calculation (shown in the figure as RS) and hash computation can start at the second cycle. Both these operations can be completed in five memory cycles. Each codeword received at the memory controller for decoding has 16 data symbols, one hash symbol and two ECC symbols. We denote the 64 data symbols and four hash symbols obtained from all the codewords which are not yet decoded by RS decoder as D' and H' respectively.  We first compute the hash (H1) of the 64 data symbols (D') and compare it with H'. 

The retrieved hash H' can match the computed hash H1 when :

\textbf{Case A1}: There is no error in H', D'  OR 
	
	\textbf{Case A2}: D' != D (the original 64-byte data stored/written in the memory) due to some error and H' = H (the original hash stored/written in the memory) but due to hash aliasing, H' = H1   OR
\\	
	\textbf{Case A3}: H' != H due to some error and D' != D due to error, but H1 (function of D') = H'. 

The retrieved hash H' does not match H1 when there is error in hash OR 64-byte data OR in both hash and 64-byte data.

In parallel, the RS decoder calculates the syndrome S$_{\textit{{\small i}}}$ for each codeword CW$_{\textit{{\small i}}}$.  S$_{\textit{{\small i}}}$ can be equal to 0 when:
\\
\textbf{Case B1}: There is no error in CW$_{\textit{{\small i}}}$   OR
\\
\textbf{Case B2}: There is an undetected error in CW$_{\textit{{\small i}}}$. 

Similarly, the syndrome is non zero when there is an error in the codeword.

Based on comparison of H1 and H' and four values of S$_{\textit{{\small i}}}$ for i=1 to 4, we come up with a decision table (Table 4). In the scenario where both the hashes match and syndrome is zero for all the four codewords (Scenario 1), we declare the cache-line to be free from errors. Theoretically, there is scope for silent data corruption here as it could be because of case A2 or A3 and case B2 for all the four codewords. From our preliminary experiments in Section 5, we can see that the probability of undetected errors is very less (0.001 \%) for each codeword. The probability reduces further when considering this scenario over all the four codewords. Therefore, we declare this scenario to be free from errors. For Scenarios 2,4 where at-least one of the syndromes S$_{\textit{{\small i}}}$ is not zero, we can check if we can correct with the help of SSC-RS and verify again with the hash. In the scenario 3, where the hashes do not match and all S$_{\textit{{\small i}}}$ are 0s, we declare the cache-line to have an undetectable error due to error in data OR in both data and hash.

As the error free scenario is more common when compared to erroneous scenarios, we design our READ operation in a way that minimises latency in the error free scenario. Therefore, as shown in Figure 6, we check for Scenario 1 at the end of five cycles and declare the cache-line to be free of errors if Scenario 1 is found to be true. Otherwise, there are two possibilities, either at-least one of S$_{\textit{{\small i}}}$ !=0 (Scenarios 2,4) OR Scenario 3. For Scenario 3, we declare the cacheline to be a uncorrectable error. In case of Scenarios 2 and 4 we use RS-SSC correction logic on codewords whose S$_{\textit{{\small i}}}$ !=0 to determine if each such codeword has a "correctable error" (CE) or "detectable-uncorrectable error" (DUE). If any one of them is an uncorrectable codeword, we declare the entire cache-line to be uncorrectable. Otherwise we correct all such codewords to obtain the corrected 64-byte data (D") and 32-bit hash (H"). In this case, there is a scope for Silent Data Corruptions (SDCs), therefore, we compute the hash H2 from D" and compare H2, H". If these hashes match, then with a high probability there is no silent data corruption. If the hashes do not match, then we can conclude that SDC occurred. 



Thus, we are able to reduce SDCs with our novel approach. On an average, the additional latency introduced per each READ miss our is expected to be one memory clock cycle. Note, that this is the similar to latency in the baseline (SSC-RS) scheme. 
\begin{table}
	
	\begin{tabular}{|c|c|c|c|}
		\hline 
		&Hash & Syndrome & Decision \\&check  &  calculation & \\ 
		\hline 
		1& H1 = H' & S$_{\textit{{\small i}}}$ = 0  & Declare Error Free \\ &&for i = 1 to 4& \\
		\hline 
		2& H1 = H'  & atleast one of  & Error, Try to correct   \\ && S$_{\textit{{\small i}}}$ != 0&it with SSC-RS and \\&&&check back with hash\\
		\hline 
		3& H1 != H' & S$_{\textit{{\small i}}}$ = 0  & Declare Error \\ &&for i = 1 to 4&\\
		\hline 
		4& H1 != H' & atleast one  & Error, Try to correct   \\ &&of S$_{\textit{{\small i}}}$ != 0&it with SSC-RS and \\&&&check back with hash\\
		\hline 
	\end{tabular} 
	\caption{Possible scenarios after initial step of Hash computation and Syndrome calculation. }
\end{table}

There is scope for false negatives (report no error although SSC decoder fails in presence of multiple symbol errors) due to hash collisions. The probability of false negative is estimated by using the upper bound on SDC rate for the baseline SSC-decoder (8\%) and collision probability for a N-bit hash is estimated by birthday paradox  $({2}^{-N/2})$. The upper bound on false negatives for our scheme is given by:
\begin{equation}
Upper \: bound \: on \: P(false \: negative) = 0.08 * {2}^{-16}
\end{equation}
\subsection{Address Protection}
Errors in address bus during memory WRITEs will lead to memory data corruption [13]. To prevent this scenario, JEDEC has introduced CAP-Command Address Parity [14] in DDR4. Another recent work, AIECC [13] proposed a stronger protection mechanism called eWRITECRC to address this concern.

With weaker CAP, errors in address bus during memory READs will result in reading codewords from an incorrect address. As the baseline CHIPKILL scheme does not keep track of address associated with the data, it will decode the codewords and inadvertently pass the data from incorrect address location to entity (I/O or processor) which initiated this READ request. Therefore, this will result in Silent Data Corruption. To provide stronger protection for up-to 32 address bits, eDECC was introduced in [13].

The 32-bit hash we used in SSCMSD design can also be used to detect multi-bit errors in the address bus during READs. We can hash all the address bits (8 bytes) along with the data during WRITE operation shown in Figure 5. This hash (H), which is stored in the form of 4 symbols in the DDRx memory will protect against both data and address (during READs) corruption. During the READ operation, as the memory controller generates the address, it already has the correct address. So, the hashes H1, H2 described in Section 6.2  will now be a function of both Data (D'/D") and the correct address. When a transmission fault results in corruption of address bits during a particular READ request (address A) [13], the memory controller in our design receives the hash and corresponding data stored in address A' (the corrupted address). At location A', we have the data and the hash of data, address (A') stored, so the RS-decoder will not be able to detect errors, but the hashes H'/H" will not be equal to H1/H2 and hence Silent data corruption is prevented.

\subsection{Hash Selection}
As described earlier, we consider non-cryptographic hash functions - Spookyhash, Lookup3 (hashlittle [38]), CRC-32 to be employed in SSCMSD.  Jenkins [15] recommends "short" version of SpookyHash for key size less than 192 bytes. We use this "short" SpookyHash for our evaluations as our key size with data, address is 72 bytes.  

Minimum hamming distance (HD) and parity are important parameters useful for deciding the generating polynomial for CRC-32. For keysize of 72 bytes, CRC-32 polynomials such as Castagnoli ({1,31}),  koopman32k ({1,3,28}), koopman32k$_{\textit{{\small 2}}}$ ({1,1,30}) provide minimum HD of 6 [39]. Therefore, errors up-to 5 (6-1) random bit flips are guaranteed to be detected by these polynomials. Errors which result in 6 or more bit-flips are not guaranteed to be detected by these polynomials. Also, the above mentioned HD=6 polynomials have even parity, hence they guarantee detection of all odd bit errors. IEEE 802.3 ({32}) polynomial provides a minimum HD of 5 for our keysize and has odd parity.

We used SpookyHash, Lookup3, CRC-32-Castagnoli (as a representative of HD=6, even parity polynomials), CRC-32-IEEE 802.3 (as a representative of odd parity polynomial) hash functions in our simulations shown in Table for SSCMSD design. Across all the fault modes, we did not find any difference in performance of  these  hash functions. Hence, we can employ one of HD=6 CRC-32 polynomials (Castagnoli / koopman32k / koopman32k$_{\textit{{\small 2}}}$)  for our SSCMSD design as they are simple, provide minimum HD=6 codewords with even parity and enable us to compute the hash in parts (due to linear property) during the READ operation. 

\begin{figure}
	\centering
	\includegraphics[scale=0.40]{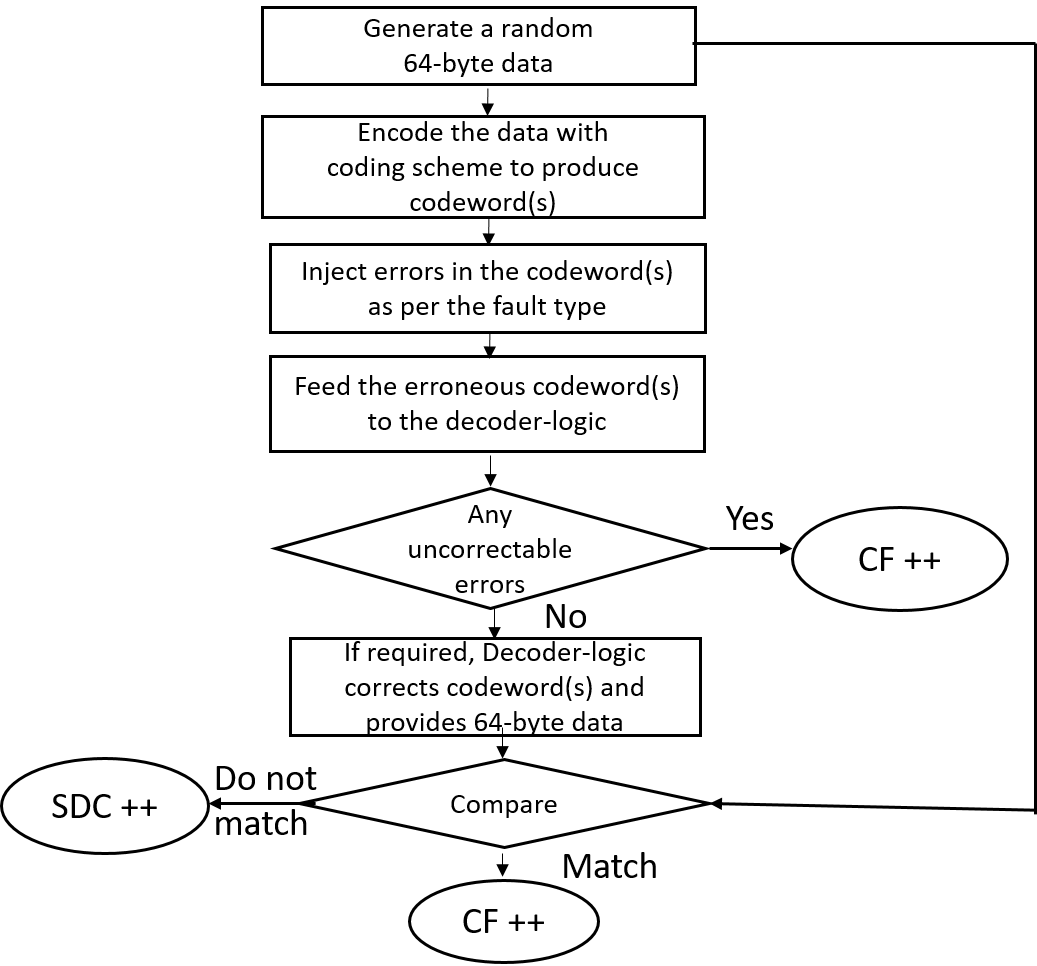}\caption{Simulation methodology.}
	
\end{figure}
\section{Evaluation}
We evaluate our scheme and compare it with the existing schemes and their extensions with the same overhead as our scheme. Baseline (RS-SSC(18,16,8)) and Bamboo-ECC (Bamboo-ECC (8 ECC symbols)) use 18 chips per rank and provide CHIPKILL capability. SSCMSD uses 19 chips per rank (Storage overhead - 18.75\%). Therefore, we extend both the baseline and Bamboo ECC with additional redundancy using their methodology of incorporating redundancy to create equal overhead conditions. RS-SSC (19,16,8) uses 3 ECC symbols, the correction capability of this code is still one symbol ($\lfloor(n-k)/2\rfloor$). The 12-ECC symbol version of Bamboo-ECC is capable of correcting up-to 6 error-symbols. 

The goal of our experiments is to compare the number of Silent Data corruptions across all the schemes for our error model. We classify fault modes to be causing up to one OR two OR three symbols/CW to be erroneous for the baseline, extended baseline and SSCMSD schemes (See Section 4). As Bamboo and extended Bamboo use vertically aligned codeword, our error model effectively translates to cause 2 to 12 symbols to be erroneous. For the rest of the discussion, we use the terminology of error model relative to the baseline scheme.

\begin{table*}[t]
	\centering
	\begin{tabular}{|c|c|c|c|c|c|c|}
		\hline 
		Comparison&Baseline  &Bamboo-ECC  &Extended Baseline   &Extended Bamboo  &SSCMSD&Stats  \\&RS(18,16,8)&(8 ECC symbols) &RS-SSC(19,16,8)&(12 ECC symbols)& RS(19,17,8) \&&\\&&&&&32-bit hash &\tabularnewline
		\hline
		
		Storage Overhead&12.5\%  &12.5\%  &18.75\%  &18.75\%  &18.75\%  &\\ECC Symbols/CW&2&8&3&12&2&\tabularnewline
		\hline
		
		Upto 1-Chip/Bus Fault&100   &100    &100    &100    &100    &CF \tabularnewline
		\hline
		Correlated & 2  &11.4  &0.9  &12.3  &0& SDC  \\Bus fault& 98&88.6&99.1&87.7&100&CF  \tabularnewline
		\hline
		1 bit fault +&4  &11.1  &2  &0  &0  & SDC\\1 bus fault&96&88.9&98&100&100&CF  \tabularnewline
		\hline
		1 bit fault + &6  &11  &3.2  &0  &0  &SDC\\(row/bank/chip)&94&89&96.8&100&100&CF \tabularnewline
		\hline 
		1 bit fault +&7.6  &0  &3  &0  &0  &SDC\\1 pin fault &92.4&100&97&100&100&CF\tabularnewline
		\hline 
		1 pin fault +&3.5  &0  &1.6  &0  &0  &SDC\\1 pin fault &96.5&100&98.4&100&100& CF \tabularnewline
		\hline 
		Chip fault +& < 0.1  &11  & < 0.1   &12  &0  &SDC\\Chip fault& > 99.9 &89& > 99.9&88&100&CF \tabularnewline
		\hline
		3 fault mode& < 0.1 &11& < 0.1&12&0&SDC\\& > 99.9 &89& > 99.9&88&100&CF \tabularnewline
		\hline
	\end{tabular} 
	\caption{Simulations results of SSC-RS, Bamboo-ECC and SSCMSD. }
\end{table*}


The following mechanisms are used to introduce errors in the encoded cacheline stored in DRAM subsystem:

1) Single bit fault :  Flip a random bit in the cacheline.

2) Single pin fault : As two beats are interleaved in the baseline scheme to form one codeword, each 8-bit symbol is composed of four 2-bit pairs. As each chip has 4 data pins, each 2-bit pair of this symbol is transferred via one pin. We therefore choose a DQ pin randomly and flip two corresponding consecutive bits of a symbol.

3) Single memory chip fault/failure : Choose a chip randomly and replace the data in the chip with a random pattern OR with all 0s OR all 1s.

4) Single bus fault: Choose a bus lane randomly and use an 8-bit random number to identify the erroneous beat positions among eight beats. As each bus-lane transfers eight beats, we then inject random errors in these positions. But, we ensure that atleast one word of this faulty bus lane is corrupted.

1-bit, 1-pin, Row/Chip/Bank, Column, Bus faults cause errors within 1 Chip or Bus lane. Correlated bus fault affects two consecutive bus lanes. As discussed in the error model, we evaluate the following 2-chip/symbol fault modes: 1 bit fault + 1 bus fault, 1 bit fault + 1 row/bank/chip fault, 1 bit fault + 1 pin fault,  1 pin fault + 1 pin fault, and chip+chip fault. 3 fault mode is also included in our evaluation.





As shown in Figure 7 for each run, we generate a 64-byte random data (representing a cache-line). The cache-line is now encoded with the specific scheme and appropriate errors are injected as per the fault mode. The corrupted encoded cacheline is fed to the corresponding decoder logic. As described earlier, Baseline, Extended-baseline and SSCMSD use four codewords per each cache-line whereas Bamboo, Extended-Bamboo use only one codeword. Accordingly, the baseline and Extended-baseline decoder logic use four RS decoders. Bamboo and extended bamboo employ only one RS decoder in their decoder logic. For SSCMSD, we use the decoder logic described in Section 6.2 (Read Operation). The decoder-logic will then determine whether this cache-line has Detectable Uncorrectable Error(s) (DUE) or  Detectable Correctable Error(s) (DCE) or no error(s). If the logic flagged any one of the codewords of the cache-line to be a DUE, we do not suspect the decoder to be wrong as our error model has multiple symbol errors (2, 3) beyond the SSC-RS correction range. In this case, the entire cacheline has to be a DUE as this cacheline cannot be consumed and we report the whole cacheline to be a correctly flagged (CF) by the decoder. For the remaining non-DUE cachelines, we compare the original (non-corrupted) 64-byte cacheline with the cumulative output of the decoder-logic. If they do not match, we report it to be a Silent Data Corruption (SDC). Otherwise we report that the scheme (decoder) correctly flagged (CF) the cacheline.

We generate one billion runs for every iteration and execute each simulation (or experiment) for 10 iterations. Table 5 lists the mean \% for these statistics across 10 iterations. The standard deviation for each of the experiments (except for SSCMSD) was up to 10,000 (for 1 billion cachelines).

As 1-bit, 1-pin, Row/Chip/Bank, Column, Bus faults result in errors confined within 1 symbol, they are corrected by all the schemes. Faults which lead to 2 or 3 symbol errors in at least one of the codeword lead to SDCs rates ranging from 0 to 7.6\% in both Baseline and extended Baseline. 1-bit + 1-pin fault  and 1-pin + 1-pin fault modes result in two symbol errors for Bamboo-ECC and extended Bamboo-ECC, hence they are corrected by them. As extended Bamboo-ECC can correct up-to six symbol errors it can provide 100\% correction with 1-bit fault + (row/chip/bank) fault and 1-bit + 1-bus fault modes. On the other hand, SSCMSD is able to avoid SDCs in all of the above fault modes. We also executed these simulations for 10 iterations, with 10 billion runs per iteration for SSCMSD scheme to understand the impact of hash aliasing. We found that there were up to 5 SDCs for each iteration across all the fault modes.  

\subsection{Experiments for Address Protection}
We also executed the simulations described in Table 5 for SSCMSD to include protection for address bits. Therefore, instead of using a random 64 bytes cache-line for each experiment (shown in Figure 7), we used random 72 bytes, each time to include 8 bytes of random address along with the cache-line. The results were identical to the ones shown in Table 5 for SSCMSD. 
In addition, we executed simulations to verify the effectiveness of SSCMSD in the presence of address errors during READs. As noted in Section 6.3, SSCMSD can provide protection against address errors during READs. Our scheme prevents SDCs due to errors in address bits during READs provided there was no address corruption during a prior WRITE operation. If one writes data to an unintended location due to address corruption, there is no way to detect such errors unless address is also stored along with data in DRAM. CAP[14] and eWRITECRC [13] can take care of address corruption during WRITEs. So, in these simulations, for each run, we generated random 72 bytes (representing the cache-line data and 8 byte address) and computed the hash (HA) of this data, address pair. Then, we used a 8-byte error mask to introduce random errors in the address bits. Next, we computed the hash (HB) of this data, corrupted address pair and compared HA and HB. If they differed, we declare that our scheme detected the errors (correctly flagged),  otherwise we declare that there was silent data corruption. We executed this simulation for 10 iterations. Each iteration comprised of 100 billion runs. Across these 10 iterations the mean of SDCs was 24.5 runs with a standard deviation of 4.3. The remaining were correctly flagged (detected as errors) by our scheme.

\section{Conclusion}

We motivate the need for addressing multiple symbol errors in CHIKPILL based DRAM subsystems given the trend of increase in failures in these systems. Based on the nature of these failures, we analyzed possible errors and then developed a new error-handling scheme called Single Symbol Correction, Multi Symbol Detection (SSCMSD). 

We implemented SSCMSD using CRC-32 and Single symbol correcting reed solomon (SSC-RS) code. By leveraging the usage of systematic SSC-RS code and simple CRC-32 hash, our novel design's impact on the READ latency is very negligible. Our simulations compare SSCMSD scheme with baseline (SSC-RS) and Bamboo-ECC. The results clearly demonstrate that SSCMSD is effective in avoiding Silent Data Corruptions (SDCs) in the presence of multiple symbol errors.

\section*{Acknowledgments}

 The research reported in this paper is partially supported by the NSF award 1618104 and the Philip and Virginia Sproul Professorship at Iowa State University. Any opinions, findings, and conclusions or recommendations expressed in this material are those of the author(s) and do not necessarily reflect the views of the funding agencies.

\def\bibfont{\footnotesize}
\bibliographystyle{IEEEtran}
\nocite{*}
\bibliography{preprint-journal}

\begin{thebibliography}{10}
\providecommand{\url}[1]{#1}
\csname url@samestyle\endcsname
\providecommand{\newblock}{\relax}
\providecommand{\bibinfo}[2]{#2}
\providecommand{\BIBentrySTDinterwordspacing}{\spaceskip=0pt\relax}
\providecommand{\BIBentryALTinterwordstretchfactor}{4}
\providecommand{\BIBentryALTinterwordspacing}{\spaceskip=\fontdimen2\font plus
\BIBentryALTinterwordstretchfactor\fontdimen3\font minus
  \fontdimen4\font\relax}
\providecommand{\BIBforeignlanguage}[2]{{%
\expandafter\ifx\csname l@#1\endcsname\relax
\typeout{** WARNING: IEEEtran.bst: No hyphenation pattern has been}%
\typeout{** loaded for the language `#1'. Using the pattern for}%
\typeout{** the default language instead.}%
\else
\language=\csname l@#1\endcsname
\fi
#2}}
\providecommand{\BIBdecl}{\relax}
\BIBdecl

\bibitem{1fb}
J.~Meza, Q.~Wu, S.~Kumar, and O.~Mutlu, ``Revisiting memory errors in
  large-scale production data centers: Analysis and modeling of new trends from
  the field.'' in \emph{DSN}, 2015, pp. 415--426.

\bibitem{2-losAlaomosdata}
\BIBentryALTinterwordspacing
Reliability data sets. Los Alamos National Laboratory. [Online]. Available:
  \url{http://institutes.lanl.gov/data/fdata/}
\BIBentrySTDinterwordspacing

\bibitem{3good}
V.~Sridharan, N.~DeBardeleben, S.~Blanchard, K.~B. Ferreira, J.~Stearley,
  J.~Shalf, and S.~Gurumurthi, ``Memory errors in modern systems: The good, the
  bad, and the ugly,'' in \emph{{ASPLOS}'15, Istanbul, Turkey, March 14-18,
  2015}, 2015, pp. 297--310.

\bibitem{4tjdell}
T.~J. Dell, ``A white paper on the benefits of chipkill-correct ecc for pc
  server main memory,'' in \emph{IBM Microelectronics Division}, 1997, pp.
  1--23.

\bibitem{5AMD-2013}
``Bios and kernel developers guide (bkdg) for amd family 15h models 00h-0fh
  processors.''\hskip 1em plus 0.5em minus 0.4em\relax AMD Inc., 2013.

\bibitem{6Jacob:2009:MSY:1855094}
B.~Jacob, \emph{The Memory System: You Can'T Avoid It, You Can'T Ignore It, You
  Can'T Fake It}.\hskip 1em plus 0.5em minus 0.4em\relax Morgan and Claypool
  Publishers, 2009.

\bibitem{7Jacob:2007:MSC:1543376}
B.~Jacob, S.~Ng, and D.~Wang, \emph{Memory Systems: Cache, DRAM, Disk}.\hskip
  1em plus 0.5em minus 0.4em\relax San Francisco, CA, USA: Morgan Kaufmann
  Publishers Inc., 2007, ch.~30.

\bibitem{8RS-nasa}
W.~A. Geisel, ``Tutorial on reed-solomon error correction coding.''\hskip 1em
  plus 0.5em minus 0.4em\relax National Aeronautics and Space Administration,
  Lyndon B. Johnson Space Center, 1990.

\bibitem{9:Udipi:2012:LLT:2337159.2337192}
A.~N. Udipi, N.~Muralimanohar, R.~Balsubramonian, A.~Davis, and N.~P. Jouppi,
  ``Lot-ecc: Localized and tiered reliability mechanisms for commodity memory
  systems,'' ser. ISCA '12, 2012.

\bibitem{10AMD-2007}
``Bios and kernel developers guide for amd npt family 15h processors.''\hskip
  1em plus 0.5em minus 0.4em\relax AMD Inc., 2007.

\bibitem{11SPARC}
``Opensparc t2 system-on-chip (soc) microarchitecture specification.'' Sun
  Microsystems, 2008.

\bibitem{12DBLP:conf/hpca/KimSE15}
J.~Kim, M.~Sullivan, and M.~Erez, ``Bamboo {ECC:} strong, safe, and flexible
  codes for reliable computer memory,'' in \emph{{HPCA}}, 2015, pp. 101--112.

\bibitem{13DBLP:conf/isca/KimSLE16}
J.~Kim, M.~Sullivan, S.~Lym, and M.~Erez, ``All-inclusive {ECC:} thorough
  end-to-end protection for reliable computer memory,'' in \emph{{ISCA}}, 2016,
  pp. 622--633.

\bibitem{14ddr4}
``Ddr4 sdram standard, jesd79-4, joint electron device engineering council,
  sep. 2012.''

\bibitem{15spookyhash}
\BIBentryALTinterwordspacing
B.~Jenkins. Spookyhash. [Online]. Available:
  \url{https://burtleburtle.net/bob/hash/spooky.html}
\BIBentrySTDinterwordspacing

\bibitem{16rscode}
\BIBentryALTinterwordspacing
H.~Minsky. A c library for reed solomon code. [Online]. Available:
  \url{http://rscode.sourceforge.net}
\BIBentrySTDinterwordspacing

\bibitem{17spookyC}
\BIBentryALTinterwordspacing
A.~Kleen. A c library for spookyhash. [Online]. Available:
  \url{https://github.com/andikleen/spooky-c}
\BIBentrySTDinterwordspacing

\bibitem{18rcode}
\BIBentryALTinterwordspacing
B.~Sklar. Reed-solomon codes. [Online]. Available:
  \url{http://ptgmedia.pearsoncmg.com/images/art_sklar7_reed-solomon/elementLinks/art_sklar7_reed-solomon.pdf}
\BIBentrySTDinterwordspacing

\bibitem{19kim2014flipping}
Y.~Kim, R.~Daly, J.~Kim, C.~Fallin, J.~H. Lee, D.~Lee, C.~Wilkerson, K.~Lai,
  and O.~Mutlu, ``Flipping bits in memory without accessing them: An
  experimental study of dram disturbance errors,'' in \emph{ACM SIGARCH
  Computer Architecture News}, vol.~42, no.~3.\hskip 1em plus 0.5em minus
  0.4em\relax IEEE Press, 2014, pp. 361--372.

\bibitem{20SSO}
\BIBentryALTinterwordspacing
Simultaneously switching noise-an overview. Mentor Graphics. [Online].
  Available:
  \url{https://www.mentor.com/pcb/blog/post/simultaneously-switching-noise-an-overview-dff75b6d-6b41-4d47-a231-1aafb29c07ad?cmpid=9049}
\BIBentrySTDinterwordspacing

\bibitem{21Kim:2012:CES:2337159.2337202}
Y.~Kim, V.~Seshadri, D.~Lee, J.~Liu, and O.~Mutlu, ``A case for exploiting
  subarray-level parallelism (salp) in dram,'' ser. ISCA '12, 2012.

\bibitem{22intelxeon}
\BIBentryALTinterwordspacing
Intel® xeon® processor e7 v2 2800/4800/8800 product family datasheet - volume
  two. Intel Corp. [Online]. Available:
  \url{http://www.intel.com/content/dam/www/public/us/en/documents/datasheets/xeon-e7-v2-datasheet-vol-2.pdf}
\BIBentrySTDinterwordspacing

\bibitem{23Memguard}
L.~Chen and Z.~Zhang, ``Memguard: A low cost and energy efficient design to
  support and enhance memory system reliability,'' ser. ISCA '14, 2014.

\bibitem{24estebanez2014performance}
C.~Est{\'e}banez, Y.~Saez, G.~Recio, and P.~Isasi, ``Performance of the most
  common non-cryptographic hash functions,'' \emph{Software: Practice and
  Experience}, vol.~44, no.~6, pp. 681--698, 2014.

\bibitem{25nelson2016ramps}
C.~Nelson, K.~R. Townsend, O.~G. Attia, P.~H. Jones, and J.~Zambreno, ``Ramps:
  A reconfigurable architecture for minimal perfect sequencing,'' \emph{IEEE
  Transactions on Parallel and Distributed Systems}, vol.~27, no.~10, pp.
  3029--3043, 2016.

\bibitem{26buserrorJacob}
B.~Jacob, S.~Ng, and D.~Wang, \emph{Memory Systems: Cache, DRAM, Disk}.\hskip
  1em plus 0.5em minus 0.4em\relax San Francisco, CA, USA: Morgan Kaufmann
  Publishers Inc., 2007, ch.~9.

\bibitem{27bus-field-study}
T.~Siddiqua, A.~Papathanasiou, A.~Biswas, and S.~Gurumurti, ``Analysis of
  memory errors from large-scale field data collection,'' in \emph{In IEEE
  Workshop on Silicon Errors in Logic - System Effects (SELSE), 2013}, 2013.

\bibitem{28bus-field-study}
V.~Sridharan and D.~Liberty, ``A study of dram failures in the field,'' in
  \emph{In International Conference on High Performance Computing, Networking,
  Storage and Analysis (SC), 2012}, 2012.

\bibitem{29soft-errors}
R.~Baumann, ``Soft errors in advanced computer systems,'' in \emph{IEEE Design
  and Test of Computers}, 2005, pp. 258--266.

\bibitem{30dramkotra}
J.~B. Kotra, N.~Shahidi, Z.~A. Chisthi, and M.~T. Kandemir, ``Hardware-software
  co-design to mitigate dram refresh overheads,'' in \emph{ASPLOS'15}, 2017.

\bibitem{31SEP}
M.~Mehrara and T.~Austin, ``Exploiting selective placement for low-cost memory
  protection,'' in \emph{ACM Transactions on Architecture and Code
  Optimization}, 2008.

\bibitem{32hashstydy}
G.~Cheng and Y.~Yan, ``Evaluation and design of non-cryptographic hash
  functions for network data stream algorithms,'' in \emph{2017 3rd
  International Conference on Big Data Computing and Communications (BIGCOM)},
  Aug 2017, pp. 239--244.

\bibitem{33Tang}
X.~Tang, M.~Kandemir, P.~Yedlapalli, and J.~Kotra, ``Improving bank-level
  parallelism for irregular applications,'' ser. MICRO-49, 2016.

\bibitem{34rsdealy}
S.~Pontarelli, P.~Reviriego, M.~Ottavi, and J.~A. Maestro, ``Low delay single
  symbol error correction codes based on reed solomon codes,'' \emph{IEEE
  Transactions on Computers}, vol.~64, no.~5, pp. 1497--1501, May 2015.

\bibitem{35incremental}
M.~v. D. B. G. G. E.~S. Dwaine~Clarke, Srinivas~Devadas, ``Incremental multiset
  hash functions and their application to memory integrity checking,'' \emph{In
  Advances in Cryptology - Asiacrypt 2003 Proceedings}, vol. 2894, pp.
  188--207, 2003.

\bibitem{36rscode}
\BIBentryALTinterwordspacing
S.~Rockliff. A c library for reed solomon code. [Online]. Available:
  \url{www.eccpage.com/rs.c}
\BIBentrySTDinterwordspacing

\bibitem{37crccombinationalLogic}
\BIBentryALTinterwordspacing
{Mytsko, Evgeniy}, {Malchukov, Andrey}, {Ryzova, Svetlana}, and {Kim, Valeriy},
  ``A study of hardware implementations of the crc computation algorithms,''
  \emph{MATEC Web of Conferences}, vol.~48, p. 04001, 2016. [Online].
  Available: \url{https://doi.org/10.1051/matecconf/20164804001}
\BIBentrySTDinterwordspacing

\bibitem{38lookup3}
\BIBentryALTinterwordspacing
B.~Jenkins. Lookup3 hash function. [Online]. Available:
  \url{https://burtleburtle.net/bob/c/lookup3.c}
\BIBentrySTDinterwordspacing

\bibitem{39Koopman02}
P.~Koopman, ``32-bit cyclic redundancy codes for internet applications,'' in
  \emph{2002 International Conference on Dependable Systems and Networks {(DSN}
  2002), 23-26 June 2002, Bethesda, MD, USA, Proceedings}, 2002, pp. 459--472.

\bibitem{40SSCMSD-PRDC}
R.~Yeleswarapu and A.~K. Somani, ``{SSCMSD} - single-symbol correction
  multi-symbol detection for {DRAM} subsystem,'' in \emph{23rd {IEEE} Pacific
  Rim International Symposium on Dependable Computing, {PRDC} 2018, Taipei,
  Taiwan, December 4-7, 2018}, 2018, pp. 15--24.

\bibitem{41euclidreedsolomon}
\BIBentryALTinterwordspacing
N.~Eruchalu. Reed-solomon (rs) encoder/decoder + channel simulation using
  euclidean algorithm. [Online]. Available:
  \url{https://github.com/nceruchalu/reed_solomon}
\BIBentrySTDinterwordspacing

\end{thebibliography}




\begin{IEEEbiographynophoto}{Ravikiran Yeleswarapu}
	 is a Ph.D candidate at Iowa State  University, Ames, Iowa. He worked at Qualcomm's WLAN division from 2010-2014. He received the bachelor's of engineering degree in electrical and electronics engineeering and the MSc degree in physics from Birla Institute of Technology and Science-Pilani, India, in 2010. His research interests include computer system design and architecture, memory, reliability and new computing paradigms.
\end{IEEEbiographynophoto}

\begin{IEEEbiographynophoto}{Arun K. Somani}
	(LF'18) received the M.S.E.E. and
	Ph.D. degrees in electrical engineering from McGill
	University, Montreal, QC, Canada, in 1983 and
	1985, respectively. He was a Scientific Officer with
	the Government of India, New Delhi, from 1974 to
	1982, and a Faculty Member with the University of
	Washington, Seattle, WA, USA, from 1985 to 1997.
	He is currently an Anson Marston Distinguished
	Professor and a Philip and Virginia Sproul Professor
	of electrical and computer engineering with Iowa
	State University.
	
	Dr. Somani's research interests are in the areas of computer system design
	and architecture, fault tolerant computing, computer interconnection networks, wavelength-division-multiplexing-based optical networking, and reconfigurable and parallel computer systems. He served as an IEEE Distinguished Visitor, an IEEE Distinguished Tutorial Speaker, and an IEEE Communication
	Society Distinguished Visitor. He delivered several keynote speeches, tutorials,
	and distinguished and invited talks all over the world. He is a Life Fellow of
	the IEEE for his contributions to theory and applications of computer networks
	in 1999. He is a Distinguished Engineer of the ACM in 2006. He is a fellow
	of the AAAS in 2012.
\end{IEEEbiographynophoto}

\end{document}